\documentclass[prb,showpacs]{revtex4}
\usepackage{graphicx,amssymb,amsmath,color,psfrag}
\begin{document}
\title{Reentrant Kondo effect in Landau quantized graphene}
\author{Bal\'azs D\'ora}
\email{dora@kapica.phy.bme.hu}
\affiliation{Max-Planck-Institut f\"ur Physik Komplexer Systeme, N\"othnitzer Str. 38, 01187 Dresden, Germany}
\author{Peter Thalmeier}
\affiliation{Max-Planck-Institut f\"ur Chemische Physik fester Stoffe, 01187 Dresden, Germany}

\date{\today}

\begin{abstract}
We have studied the interplay of an Anderson impurity in Landau quantized graphene, with special emphasis on the 
influence of the chemical potential. Within the slave-boson mean-field theory, we found reentrant Kondo behaviour by 
varying the chemical potential or gate voltage. Between Landau levels, the density of states is suppressed, and by 
changing the graphene's Fermi energy, we cross from  metallic to semiconducting regions. Hence, the corresponding Kondo 
behaviour is also influenced.
The f-level spectral function reveals both the presence of Landau levels in the conduction band and the Kondo 
resonance.
\end{abstract}

\pacs{81.05.Uw,71.10.-w,73.43.Qt}

\maketitle

\section{Introduction}



The existence of Dirac fermionic excitations  in single layer graphene has attracted great 
interest\cite{berger,novoselov1,novoselov3,bostwick,zhou,geim}
Theoretically the linear dispersion and chiral nature of massless quasiparticles has many unexpected 
physical consequences, especially for magnetotransport  properties\cite{peresalap,ando1,sharapov3,ziegler,doragraph}. 
Experimentally graphene may be 
considered as a zero gap semiconductor with charge carriers of very high mobility whose density may 
easily be controlled by applying a gate voltage which gives hope for interesting applications \cite{geim}.

The scattering of graphene quasiparticles from normal impurities determines electronic and thermal 
transport and is well studied \cite{peresalap,sharapov3,ziegler,doragraph,Loefwander07,adam}. The linear dispersion of 
quasiparticles and the 
associated pseudogap plays an essential role in the magnetotransport \cite{doragraph}  because it leads 
to unconventional Landau level quantization as compared to the case of parabolic bands. One essential 
difference is that a zeroth order Landau level is pinned at zero energy for any field strength. With 
increasing impurity scattering the pseudogap is therefore gradually filled and the density of states 
exhibits oscillations as function of the chemical potential or carrier number\cite{sharapov3,doragraph}. 

Because the graphene carrier number and density of states can easily be manipulated with gate voltage and 
magnetic field  it is attractive to consider the effect of magnetic impurities in a graphene host. It is 
well known that magnetic moments in the Fermi sea show the competition of on-site Kondo singlet formation 
and intersite RKKY interactions which is highly sensitive to the density of states around the Fermi 
level\cite{vozmediano}. 
While experiments in graphene are still missing we think that chemisorption or adsorption of transition metal ions on 
graphene sheets is the most likely realization of Kondo effect and RKKY interaction with Dirac electrons. 
Possible ways for creation of local moments  in carbon based materials have been discussed in Refs. 
\onlinecite{harigaya,sengupta,Brey07}.

Theoretically the Kondo and Anderson- type models for local moments in graphene have been studied without 
magnetic field \cite{sengupta,hentschel}. In the present work we investigate the 
effect of Landau level formation on 
the 
screening of magnetic  impurities described by the Anderson model.  Due to the strong DOS variation with 
chemical potential and magnetic field it should be possible to drive the magnetic impurity in and out of 
the Kondo regime which is characterized by the formation of an f-level resonance close to the chemical 
potential. This effect is studied within mean field slave boson theory and assuming a constant broadening 
of Landau levels by normal impurities. In Sect. II we define the Anderson impurity model coupled to a Dirac 
Fermi sea in a magnetic field. The solution of the model within the saddle point approximation is derived in 
Sect. III and the numerical results are discussed in Sect. IV. Finally Sect. V gives the conclusion.

\section{Anderson impurity in graphene}

The Hamiltonian of Dirac quasiparticles living on a single graphene sheet, interacting with an infinite-$U$ Anderson 
impurity at the origin is given by\cite{semenoff,gonzalez,peresalap}:
\begin{gather}
H_0=\int d{\bf r}\sum_{\sigma,s}\Psi^+_{\sigma,s}({\bf 
r})\left[\sum_{j=x,y}-v_F(1-2\delta_{j,y}\delta_{s,-1})\sigma_j\left(-i\partial_j+eA_j(\bf 
r)\right)-\mu-h\sigma\right] \Psi_{\sigma,s}({\bf r})+\sum_\sigma (E-h\sigma) f^+_\sigma f_\sigma+\nonumber\\
+V\sum_{\sigma,s}\left(f^+_\sigma\Psi_{\sigma,s}({\bf 0})b+b^+\Psi^+_{\sigma,s}({\bf 
0})f_\sigma\right) ,
\label{hamalap}
\end{gather}
where $\sigma_j$'s are the Pauli matrices, and stand for Bloch states residing on the two different sublattices of the 
bipartite hexagonal lattice of graphene\cite{peresalap,sharapov3}.
The quasiparticle spectrum vanishes at six point in the Brillouin zone. Out of these six, only two are non-equivalent 
points,
and are referred to as  $K$ and $K^\prime$ points in the Brillouin zone, denoted by 
the $s=1$ and -1 index, respectively, $\delta$ is the Kronecker delta. The Kronecker delta function accounts for 
the non-equivalence of the two 
Dirac points\cite{ando1} at $K$ and $K^\prime$.
The finite chemical potential $\mu$ accounts for particle-hole symmetry breaking. 
The 
vector potential for a constant, arbitrarily oriented  magnetic 
field reads as ${\bf A(r)}=(-By\cos\theta,0,B(y\sin\theta\cos\phi-x\sin\theta\sin\phi))$, where $\theta$ is the 
angle the magnetic field makes from the $z$ axis, and $\phi$ is the in-plane polar-angle measured from the $x$-axis.
The Zeeman term is assumed to couple to both the impurity and Dirac electrons by the same $g$ factor, $h=g\mu_B B$, 
$v_F\approx 10^6$~m/s, is characteristic to graphene. 
 $f^+_\sigma$ and $f_\sigma$ creates and annihilates an electron on the localized $E$ level,
$b^+$ and $b$ are the slave boson operators, responsible for the hole states\cite{bickers,coleman,hewson}.
These take the infinite-$U$ term into account.

The Hamiltonian should be restricted to the subspace
\begin{equation}
\sum_\sigma f^+_\sigma f_\sigma+b^+b=1. 
\label{constr}
\end{equation}

Within the mean-field approximation, the slave-boson operators are replaced by their expectation value, $b_0=\langle b
\rangle$, and the constraint is
satisfied by introducing a Lagrange multiplier $\lambda$:
\begin{gather}
H=\int d{\bf r}\sum_{\sigma,s}\Psi^+_{\sigma,s}({\bf 
r})\left[\sum_{j=x,y}-v_F(1-2\delta_{j,y}\delta_{s,-1})\sigma_j\left(-i\partial_j+eA_j(\bf
r)\right)-\mu-h\sigma\right] \Psi_{\sigma,s}({\bf r})+\sum_\sigma (E+\lambda-h\sigma) f^+_\sigma f_\sigma+\nonumber\\
+Vb_0\sum_{\sigma,s}\left(f^+_\sigma\Psi_{\sigma,s}({\bf 0})+\Psi^+_{\sigma,s}({\bf 0})f_\sigma\right)+\lambda(b_0^2-1).
\label{mfh}
\end{gather}
In the particle-hole symmetric case ($\mu=0$), in the absence of magnetic field, the energy spectrum of the 
system of Dirac electrons is 
given by 
\begin{equation}
E({\bf k})=\pm v_F|\bf k|.
\end{equation}
This describes massless relativistic fermions with spectrum consisting of two cones, touching each other at the 
endpoints.
From this, the density of states per spin follows as
\begin{equation}
\rho(\omega)=\frac 1\pi \sum_{\bf k}\delta(\omega-E({\bf k}))=\frac 1\pi 
\frac{A_c}{2\pi}\int\limits_0^{k_c}kdk\delta(\omega\pm v_F k)=\frac{2|\omega|}{D^2},
\end{equation}
where $k_c$ is the cutoff, $D=v_Fk_c$ is the bandwidth, and $A_c=4\pi/k_c^2$ is the area of the hexagonal unit cell.
We mention in passing, that an applied gate voltage directly controls the number of extra charge carriers in the 
system, which is given by $eV\sim n=\int_0^\mu\rho(\omega)d\omega=\mu^2/D^2$. Hence, chemical potential is proportional 
to 
the 
square root of the applied gate voltage even in the ideal case, without any scatterers and magnetic field.
Such a relation can hardly be calculated for the realistic case. Nevertheless, the chemical potential always varies 
monotonically with the gate voltage due to the positiveness of the density of states.

Magnetic impurities in gapless fermi systems have thoroughly been studied starting with the pioneering work 
of Withoff and Fradkin\cite{WF}, and 
the focus was on the influence of gapless excitations on the Kondo phenomenon (for a review, see Ref. 
\onlinecite{balatsky}). Recently, Kondo effect in graphene 
without magnetic field has been studied within this framework\cite{sengupta}.
Here we allow for Landau quantization of 
the quasiparticle spectrum, and study the orbital effect of magnetic field on the various Kondo phases.

In the presence of magnetic field, the eigenvalue problem of our Hamiltonian without the 
localized 
level can readily be 
solved\cite{peresalap}. 
From now on, we concentrate on the $K$ point, the eigenfunctions of the $K^\prime$ point can be 
obtained by exchanging the two components of the spinor. Momentarily, we also neglect the spin and the Zeeman term, and 
concentrate 
on the effect of Landau quantization. They will be reinserted when necessary.
For the zero energy mode (E=0), the 
eigenfunction is obtained as
\begin{gather}
\Psi_k({\bf r})=\frac{e^{ikx}}{\sqrt{L}}\left(\begin{array}{c}
0 \\
\phi_0(y-kl_B^2)
\end{array}\right),
\end{gather} 
and the two components of the spinor describe the two bands. The energy of the other modes reads as
\begin{gather}
E(n,\alpha)=\alpha\omega_c\sqrt{n+1}
\label{landauenergia}
\end{gather}
with $\alpha=\pm 1$, $n=0$, 1, 2,\dots, $\omega_c=v_F\sqrt{2e|B\cos(\theta)|}$ is the Landau scale or energy, but is 
different 
from the cyclotron frequency\cite{zheng}. 
Only the perpendicular component of the field enters into these expressions, and by tilting the field away from the 
perpendicular direction corresponds to a smaller effective field.
The sum over integer $n$'s is cut off  
at $N$ given by $N+1=(D/\omega_c)^2$, which means that we consider $2N+3$ Landau levels altogether. 

The corresponding wave function is
\begin{gather}
\Psi_{n,k,\alpha}({\bf r})=\frac{e^{ikx}}{\sqrt{2L}}\left(\begin{array}{c}
\phi_n(y-kl_B^2) \\
\alpha\phi_{n+1}(y-kl_B^2)
\end{array}\right)
\end{gather}  
with cyclotron length $l_b=1/\sqrt{eB}$. Here $\phi_n(x)$ is the $n$th eigenfunction of the usual one-dimensional 
harmonic oscillator. The electron-field operator at the $K$ point can be built up from these functions as
\begin{gather}
\Psi({\bf r})=\sum_k\left[ \Psi_k({\bf r})c_{k}+\sum_{n,\alpha}\Psi_{n,k,\alpha}c_{k,n,\alpha}\right].
\end{gather}
The Green's functions of these new operators do not depend on $k$, and read as
\begin{gather}
G_0(i\omega_n,k)=\frac{1}{i\omega_n},\\
G_0(i\omega_n,k,n,\alpha)=\frac{1}{i\omega_n-E(n,\alpha)}
\end{gather}
for $c_k$ and $c_{k,n,\alpha}$, respectively, and $\omega_n$ is the fermionic Matsubara frequency.
As seen from above, the density of states in the presence of quantizing magnetic field contains Dirac-delta peaks 
located at the Landau level energies, and quasiparticle excitations have infinite lifetime. To describe a more realistic 
situation, scattering from disorder needs to be considered in the presence of the magnetic field. Usually, the resulting 
self energy of the Dirac fermions, determined in a self-consistent manner, 
depends on the frequency and field strength\cite{peresalap,doragraph}. However, good 
agreement can be reached by assuming a constant, phenomenological scattering rate, denoted by $\Gamma$, for small and 
moderate fields, as can be learned from similar analyzis\cite{sharapov2,sharapov3}. To simplify calculations, we  
have chosen to mimic disorder by a constant scattering rate.

\section{Saddle-point equations}

The free energy of the system can be found from Eq. \eqref{mfh} using standard technique\cite{hewson}. 
The value of $\lambda$ and $b_0$ is determined self-consistently by minimizing the free energy of the system with 
respect to 
them\cite{zhang,zhu}. As a
result, by restoring the finite chemical potential, spin and the Zeeman term, the
saddle point equations at $T=0$ are given by:
\begin{gather}
b_0^2=\sum_{\sigma=\pm 1}S_1(E+\lambda-\sigma h,\mu+\sigma h),\label{speb}\\
\lambda=\sum_{\sigma=\pm 1}S_2(E+\lambda-\sigma h,\mu+\sigma h)\label{spel},
\end{gather}
where the auxiliary functions are defined as
\begin{gather}
S_1(y,z)=\int\limits_0^\infty \frac{dx}{\pi} 
\frac{y+b_0^22V^2\textmd{Im}\Sigma(x+\Gamma-iz)}
{(x+b_0^22V^2\textmd{Re}\Sigma(x+\Gamma-iz))^2+(y+b_0^22V^2\textmd{Im}\Sigma(x+\Gamma-iz))^2},\\
S_2(y,z)=2V^2\int\limits_0^\infty 
\frac{dx}{\pi}\frac{(x+b_0^22V^2\textmd{Re}\Sigma(x+\Gamma-iz))\textmd{Re}\Sigma(x+\Gamma-iz)+\textmd{Im}\Sigma(x+\Gamma-iz)
(y+b_0^22V^2\textmd{Im}\Sigma(x+\Gamma-iz))}
{(x+b_0^22V^2\textmd{Re}\Sigma(x+\Gamma-iz))^2+(y+b_0^22V^2\textmd{Im}\Sigma(x+\Gamma-iz))^2},\\
\Sigma(z)=\frac{1}{N+1}\left[\frac{1}{z}+\sum_{k=0}^N 
\frac{2z}{z^2+\omega_c^2(k+1)}\right]=\frac{1}{N+1}\left[\frac{1}{z}+2\frac{z}{\omega_c^2}\left\{
\Psi\left(\frac{z^2}{\omega_c^2}+ N + 
2\right)-\Psi\left(\frac{z^2}{\omega_c^2}+1\right)\right\}\right].
\end{gather}
The $1/(N+1)\propto B$ prefactor denotes the Landau level degeneracy. These reduce to the standard saddle-point 
equations for gapless phases\cite{zhang,doraanderson} for zero field and $\Gamma=0$.
The extra factor of 2 in front of $V^2$ stems from the two non-equivalent Dirac cones at the $K$ and $K^\prime$ points, 
since by using
\begin{equation}
\int\limits_{-\infty}^\infty dk\phi_n(y-kl_B^2)\phi_m(y-kl_B^2)=\frac{\delta_{n,m}}{l_B^2},
\end{equation}
each Landau level at each valley contributes to the hybridization energy by $V^2$. 
Eq. \eqref{speb} accounts for the constraint of having at most one f-electron at the impurity site (Eq. \eqref{constr}), 
Eq. \eqref{spel} stems from the equation of motion of the slave boson field $b$: since it is constant in the mean-field 
approach, the terms determining its dynamics should add up to zero\cite{coleman}.
Here $\Sigma(z)$, which is related to the f-level self energy, contains all the information about the conduction 
electron bath, where the magnetic impurity 
is 
embedded. When the field strength goes to zero, these equations reduce to those found in gapless phases\cite{zhang} 
as
\begin{equation}
\Sigma(z)=2\frac{z}{D^2}\ln\left(1+\frac{D^2}{z^2}\right).
\end{equation}
In this case, for $\mu=\Gamma=0$, the critical f-level energy is found to be $E_c=-8V^2/D$. For $E<E_c$, only the trivial 
solution occurs ($b_0=0$), hence charge fluctuations are completely suppressed.
The solution of Eqs. \eqref{speb}-\eqref{spel} can be carried out by realizing, that Eq. \eqref{speb} depends only on 
the 
renormalized f-level energy $E+\lambda$, and not separately on the two variables $E$ and $\lambda$. Then, by fixing the 
value of 
$E+\lambda$, we can determine the 
corresponding $b_0$ by iteration, for example. By inserting the values of the renormalized f-level energy and 
the 
slave-boson expectation value to Eq. \eqref{spel}, we can directly read off $\lambda$, and determine $E$ by subtracting 
it from the renormalized f-level energy. As in other similar problems, this method predict a quantum phase transition at 
$T=0$, characteristic to gapless Kondo phases. However, in our case, the order of the transition can change from second 
to first. Such a transition is absent for magnetic impurities embedded to normal metals\cite{hewson}.

\section{Spectral function, discussion}

\begin{figure}[h!]
\includegraphics[width=8cm,height=8cm]{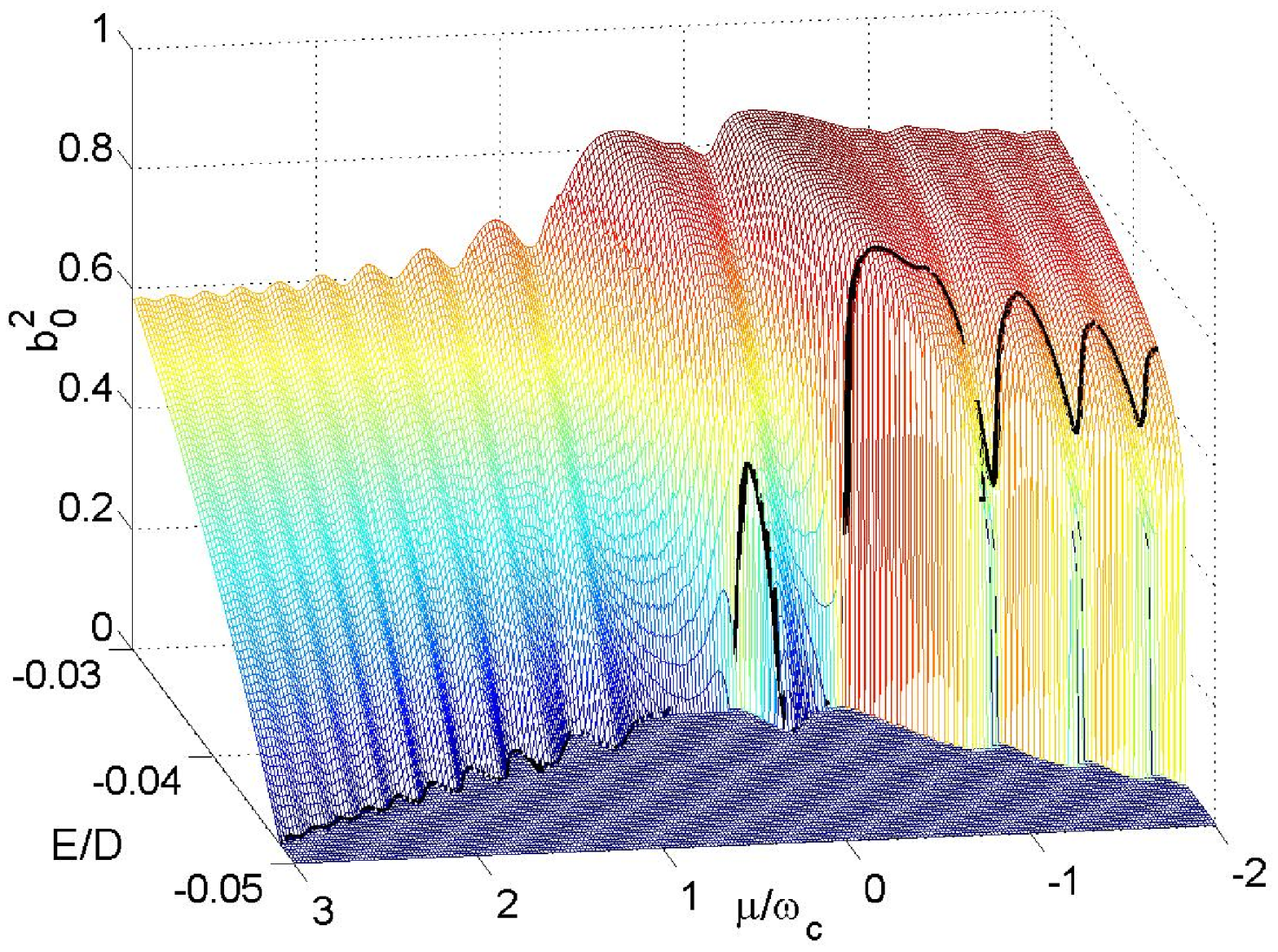}
\hspace*{5mm}
\includegraphics[width=8cm,height=8cm]{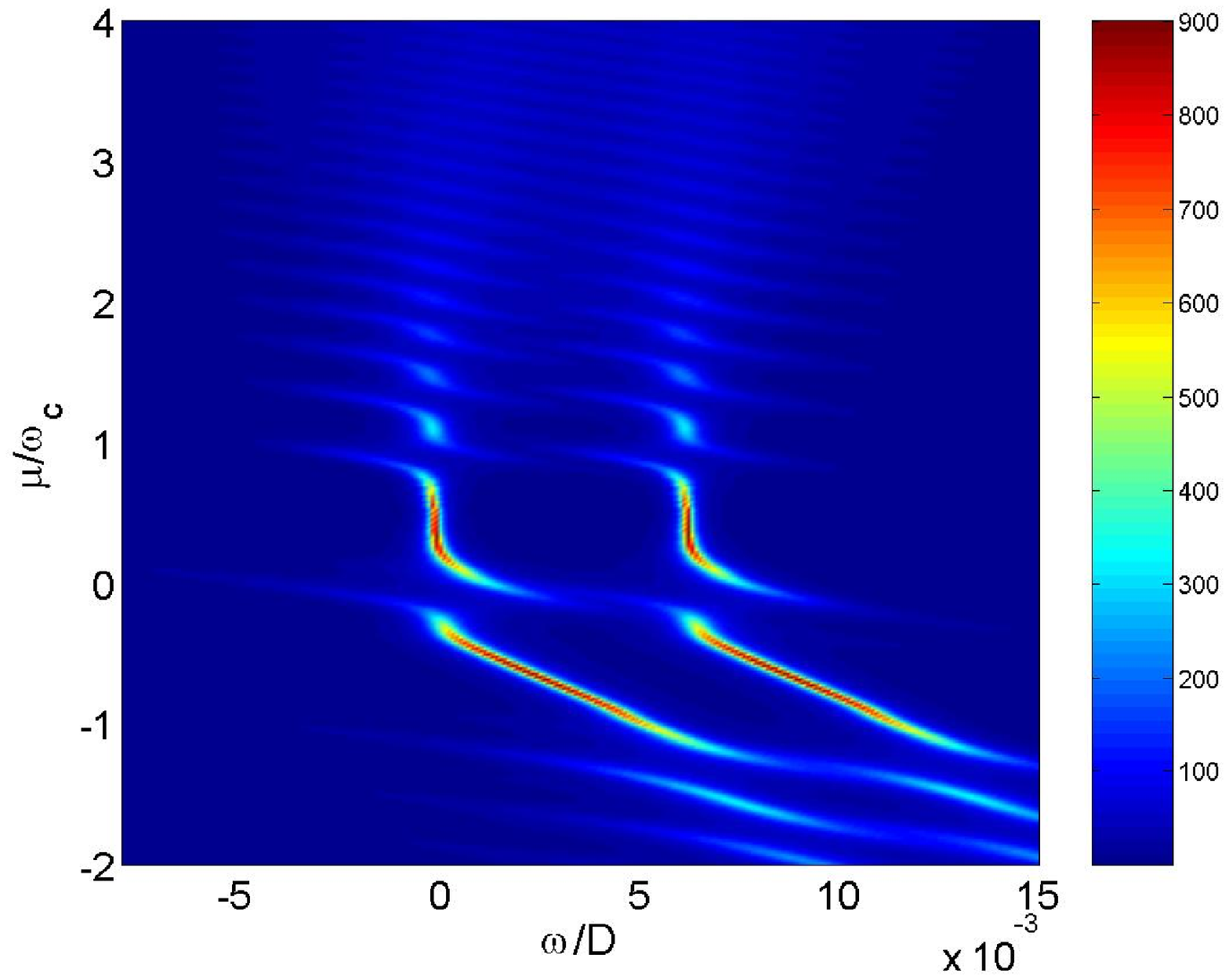}
\caption{(Color online) The order parameter (left panel) and the contour plot (for $E/D$=-0.036, right panel) of the 
f-level density of states 
are 
shown for $N=1000$, $2(V/D)^2=0.01$, $\Gamma=0.05\omega_c$, $h=0.1\omega_c$. 
For negative chemical potentials, the order of the transition changes from second to  first, as is denoted by the thick 
black line. 
The parallel structures in the contour plot denote the Kondo peaks, separated by twice the Zeeman energy.}
\label{ordpar}
\end{figure}

\begin{figure}[h!]
\includegraphics[width=8cm,height=8cm]{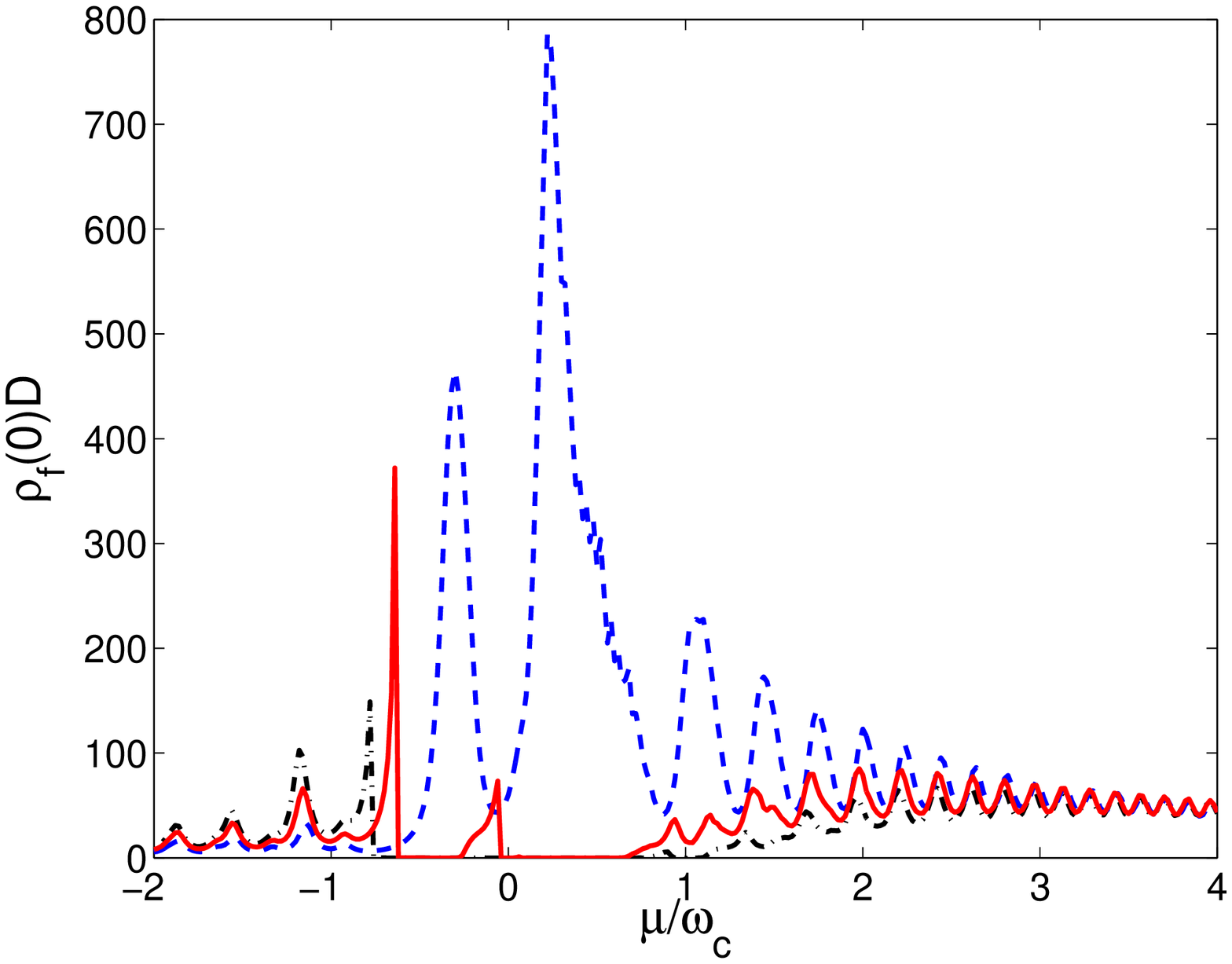}
\hspace*{5mm}
\includegraphics[width=8cm,height=8cm]{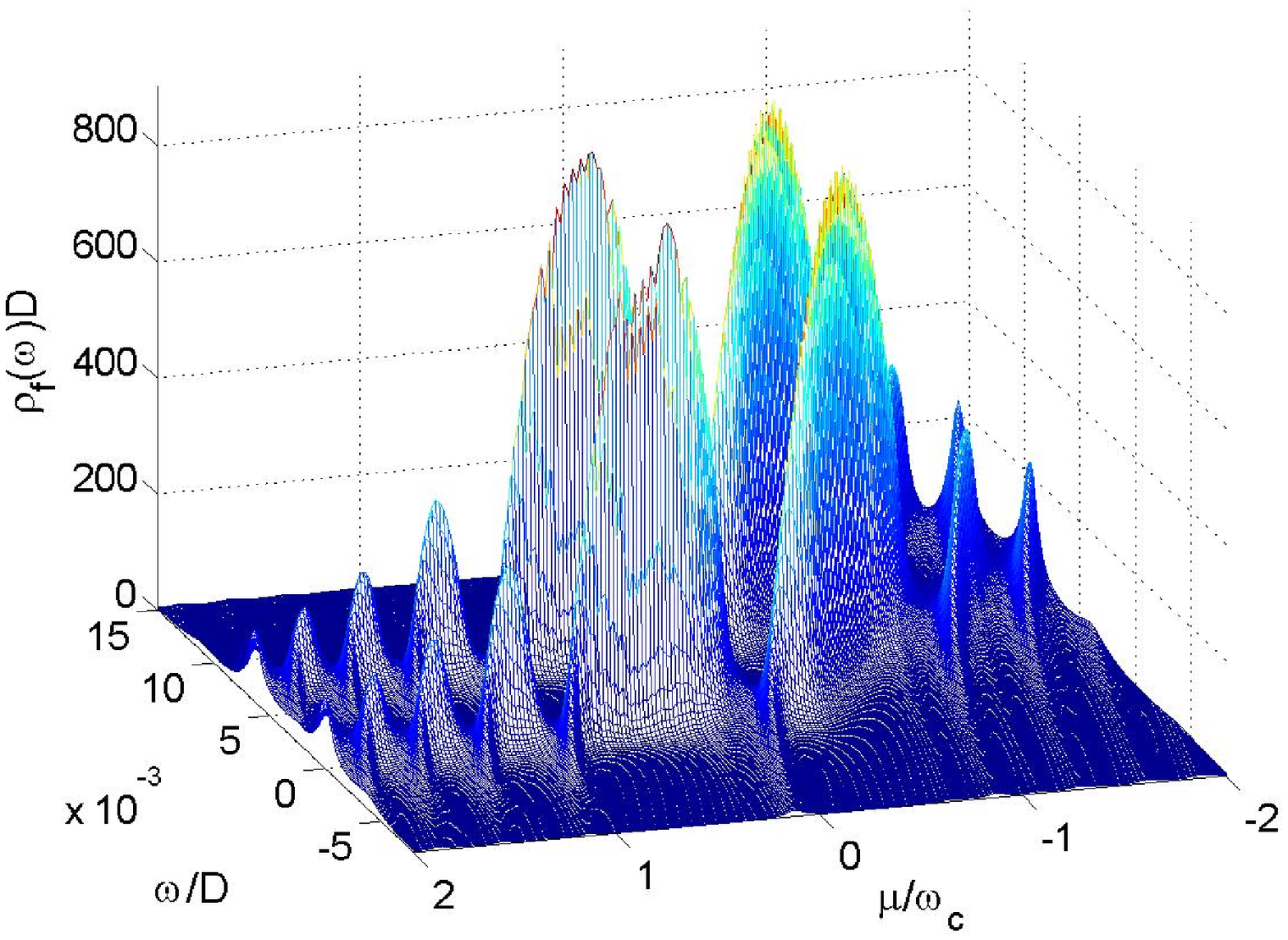}
\includegraphics[width=8cm,height=8cm]{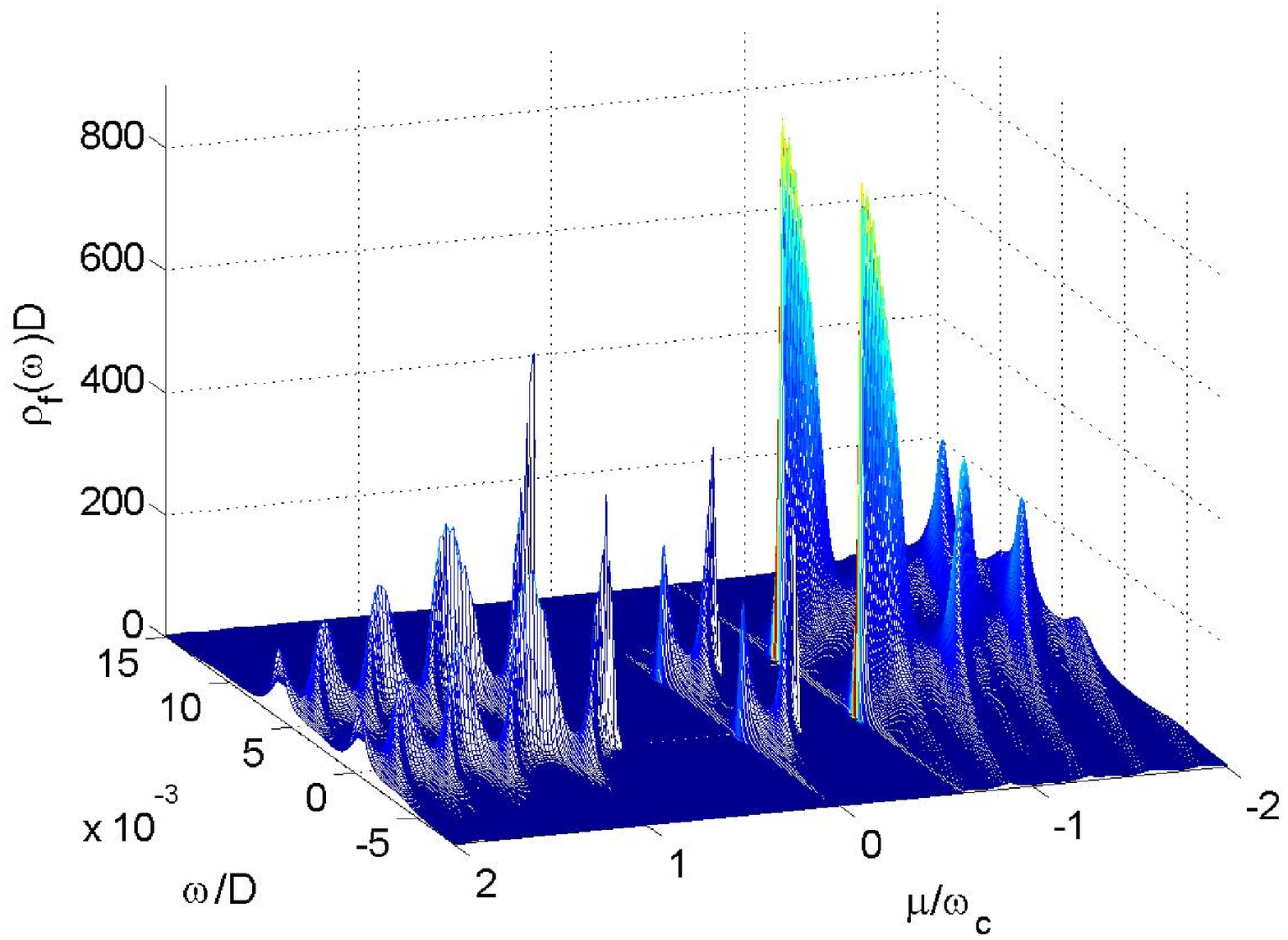}
\hspace*{5mm}
\includegraphics[width=8cm,height=8cm]{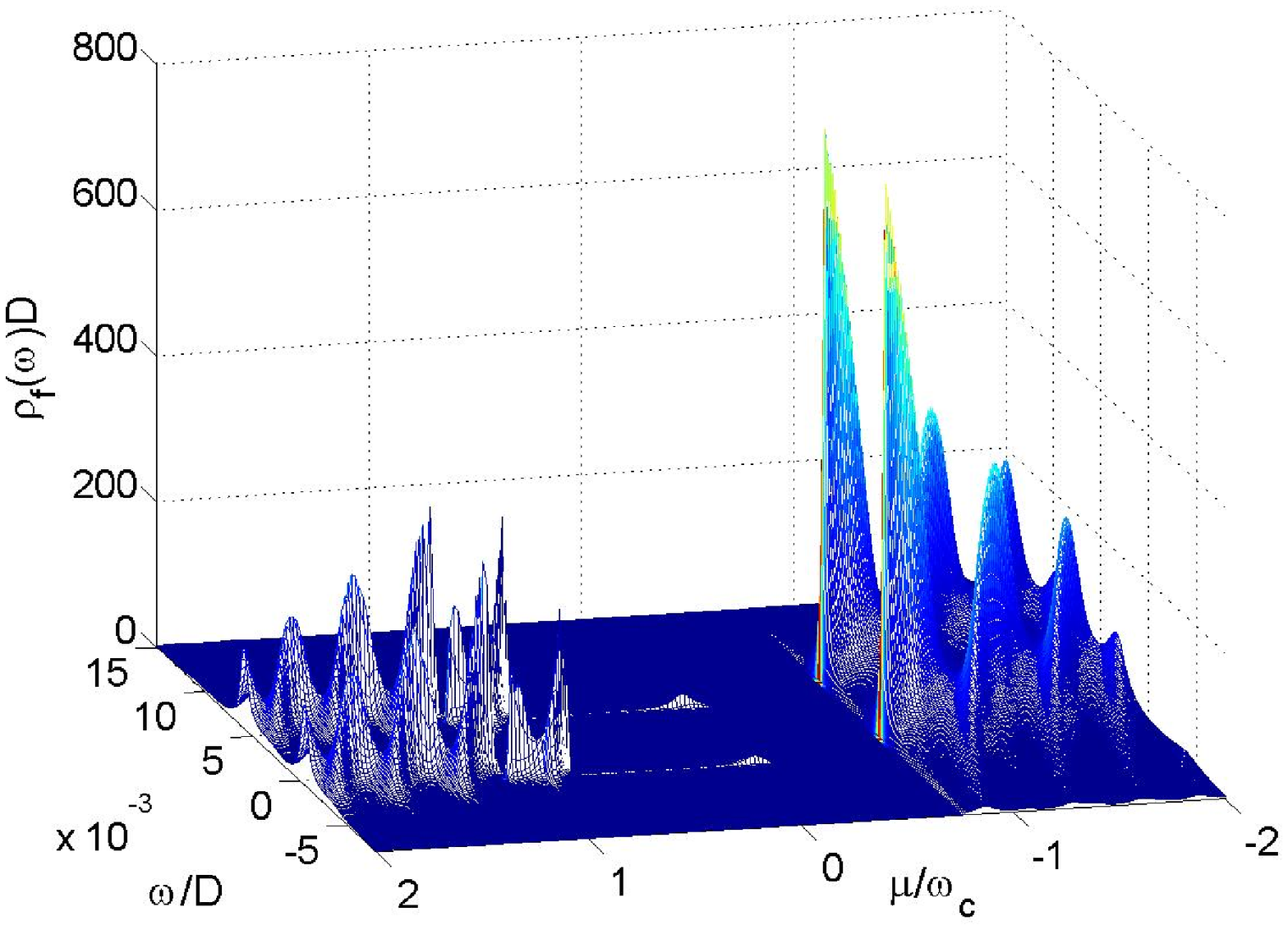}
\caption{(Color online) In the top left panel, the spin averaged f-level density of states is shown as a function of 
the chemical potential  
for
$N=1000$, $2(V/D)^2=0.01$, $\Gamma=0.05\omega_c$, $h=0.1\omega_c$ for $E/D$=-0.036 
(blue dashed), -0.0387 (red solid) and -0.04 (black dashed dotted). Note the presence of states 
for the middle value close to zero 
chemical potential. Reentrant behaviour is also observable close to the first Landau level for $E/D$=-0.04, 
$\mu\approx\omega_c$.
The three 3 dimensional plots show the evolution of the spectral density as a function of chemical potential and 
frequency for the previous three $E$ values from top to bottom, left to right. Note the presence of small island of 
states for $E/D=-0.0387$, responsible for the Kondo effect at $\mu\approx 0$.}
\label{dos3d}
\end{figure}

The full f-electron spectral function per spin along the real frequency axis reads as
\begin{gather}
\rho_{f,\sigma}(\omega)=-\frac 1\pi \textmd{Im}\frac{b_0^2}{\omega-E-\lambda+\sigma 
h-2V^2b_0^2\Sigma_f(\omega+\mu+\sigma h+i\Gamma)},
\end{gather}
where $\Sigma_f(x)=-i\Sigma(-ix)$.
It shows the Landau level oscillations, and in addition, a large Kondo peak whenever non-trivial solution of Eqs. 
\eqref{speb}-\eqref{spel} exists. 
Without magnetic field and normal impurities, it shows marginal Fermi liquid behaviour\cite{varma}, as can be observed 
from the 
analytically
continued f-level self-energy:
\begin{equation}
\Sigma_f(\omega \ll D)\approx -\frac{4V^2b_0^2}{D^2}\left[2\omega\ln\left(\frac{D}{|\omega|}\right)+i\pi 
|\omega|\right], 
\end{equation}
which, in the presence of normal impurities, turns into the usual fermionic self energy at low energies, with $\omega$ 
being replaced by $\Gamma$, the scattering rate.
In the case of quantizing magnetic field, without normal impurities, its imaginary part consists of Dirac-delta 
functions at the Landau level energies. Non-magnetic impurities smear these singularities, by transforming them
 into Lorentz functions. Hence, the self energy is that of a well-formed Fermi liquid for realistic 
situations. We mention the possibility of detecting non-Fermi liquid phases\cite{sengupta,harigaya} in graphene due to 
the valley degeneracy, which might originate from multichannel Kondo physics. This can be signaled by the 
finite critical value of $E$ even at $T=0$ as is seen in Fig. \ref{ordpar}. 
However, such a situation is unlikely to be observed by our simple mean-field analysis.

The numerical solution of the saddle-point equations have been carried out, and the result for the order parameter 
($b_0^2$) is shown in Fig. \ref{ordpar}. It is directly related to the f-level occupation through Eq. 
\eqref{constr} 
as $n_f=1-b_0^2$. 
In general, by increasing $|\mu|$, a Fermi surface develops, and the critical value of $E$ penetrates deeply into the 
$E<0$ region\cite{WF}. However, there is a crucial difference between positive and negative chemical potentials: for the 
former, 
the Kondo energy scale becomes extremely small, as can be seen in Fig. \ref{ordpar} in the f-level density of states, 
since the f-level energy is well below the Fermi energy of conduction electrons. In the latter case, the 
energy level of f-electrons lies above the Fermi energy, hence the Kondo scale enhances.

When the chemical potential is close to a Landau level energy $E(n,\alpha)$, the non-trivial solution (nonzero 
$b_0$) extends further 
in the $E<0$ region, because the density of states is enhanced around Landau level energies.
Between Landau levels, the conduction electron density of states is suppressed, and the critical f-level energy moves 
closer to zero, leading to oscillatory behaviour in the phase boundary and diagram, as can be seen in Fig. \ref{ordpar}. 
Here we assume $N=1000$, which corresponds to weak or moderate fields\cite{doragraph}, depending on the explicit value 
of the cutoff $D$. The Zeeman term is chosen to be much smaller than $\omega_c$, is follows from actual numbers in 
graphene\cite{sharapov3}.
As a result, by changing the chemical potential, we can move between Landau levels, and we can enter into and leave the 
Kondo regime. Hence reentrant behaviour is found. By decreasing the scattering rate $\Gamma$ from normal impurities, the 
oscillations along the 
phase boundary become more pronounced.

The reentrant behaviour can more directly be checked in the f-level density of 
states in 
Fig. \ref{dos3d}, 
where large Kondo peaks are observable (due to Zeeman splitting) in addition to small oscillation stemming from Landau 
levels in the conduction band, when non-trivial solution exists. 
The distance between the two parallel ridges in the contour plot in Fig. \ref{ordpar} is two times the Zeeman energy, as 
it should be.
When only the trivial solution exists ($b_0=0$), $\rho_f(\omega)$ is 
completely suppressed, since the maximally allowed one particle always occupies the f-level.

In the Kondo regime, as one varies the chemical potential, the Kondo temperature does not change monotonically. It 
remains mainly pinned to the closest Landau level, and then suddenly jumps to the neighbouring one, as is seen in Fig. 
\ref{ordpar}.

These features in the density of states can probably be detected by conductance measurements, which measures directly 
the 
inverse of the f-level density of states, in addition to normal impurities\cite{hewson}. 
When $\rho_f(0)$ is finite, its contribution 
is thought to overwhelm that of normal impurities\cite{hewson}. Then, both the Landau level oscillation stemming from 
the orbital quantization of conduction electrons and the Kondo behaviour could be seen. 
The change in the Kondo 
temperature (the peak position on Fig. \ref{ordpar} and \ref{dos3d}) by chemical potential or gate voltage can also in 
principle be 
detected from the specific heat. This is expected to exhibit a double peak 
structure around $T_K$ due to the Zeeman 
term. However, such measurement on thin graphene films are extremely demanding. 
The change of the spectral function would reveal itself directly in photoemission spectroscopy\cite{zhou,bostwick}, 
which, however, in a 
finite magnetic field, does not constitute a standard experiment.

The presence of Kondo resonance makes itself felt in magnetic responses, which are expected to be different 
from 
the usual Kondo behaviour due to the presence of orbital (or diamagnetic) terms. This already influences the magnetic 
susceptibility of pure graphene, and probably alters the magnetic behaviour of the impurity as well.

\section{Conclusions}

In summary, we have studied the infinite-U Anderson impurity embedded to a host of two-dimensional Dirac fermions 
within the self-consistent slave boson mean field theory. The host material corresponds to graphene, where the 
elementary excitations on a honeycomb lattice are Dirac fermions. Such a system can most probably be 
realized by the chemisorption of transition metal ions on graphene sheets.
We allow for the Landau quantization of the conduction electron spectrum, which turns out to be 
unusual\cite{peresalap,andrei} 
($E\propto\sqrt{n}$) in comparison with normal metals ($E\propto {n}$). Then, we study the effect of 
\textit{orbitally 
quantizing magnetic field} on the Kondo phenomenon, in addition to the Zeeman term, albeit the latter is thought to be 
negligible in the presence of Landau levels\cite{sharapov3}. 

When the chemical potential lies close to a Landau level energy, the conduction electron density of states is 
enhanced, and the 
mixed-valence regime extends further in the $E<0$ region\cite{WF}. Between Landau levels, the density of states 
resembles to that in an insulator, and the local moment regime gains ground. Hence by varying the chemical 
potential between Landau levels, 
reentrant Kondo behaviour is found. This manifests itself in the f-electron spectral function, which accommodates small 
islands of states, corresponding to the mixed-valence case, separated by the deserts of local moment region, as $\mu$ 
changes. 
The reentrant behaviour manifest itself strongly around the zeroth order Landau level, but for clean samples, it should 
be observable at higher levels as well.
 
The experimental detection of this phenomenon can be done by conductance measurements in magnetic field at low 
temperatures, which can show the presence or absence of Kondo peaks in the f-electron density of states.

\begin{acknowledgments}

This work was supported by the Hungarian Scientific Research Fund under grant number OTKA TS049881. 
\end{acknowledgments}

\bibliographystyle{apsrev}
\bibliography{refgraph,anderson}
\end{document}